\documentclass[12pt]{article}
\usepackage[colorlinks=true,linkcolor=blue,urlcolor=blue,filecolor=black,citecolor=red,pdfstartview=FitV,pdftitle={},pdfsubject={},pdfkeywords={},pdfpagemode=None,bookmarksopen=true]{hyperref}
\usepackage{graphicx}
\usepackage{epstopdf}%
\usepackage{amsmath}
\usepackage{amsfonts}
\usepackage{amssymb}
\usepackage{color}%
\usepackage{dcolumn}
\usepackage{slashed}
\usepackage{amssymb,ulem}
\usepackage{float}
\usepackage{amsthm,amsmath,amssymb}
\usepackage{mathrsfs}
\usepackage{subfigure}
\usepackage{wrapfig}
\providecommand{\U}[1]{\protect\rule{.1in}{.1in}}

\newcommand{\f}{\begin{equation}}
\newcommand{\ff}{\end{equation}}
\newcommand{\fa}{\begin{eqnarray}}
\newcommand{\ffa}{\end{eqnarray}}

\begin{document}
\title{Joule-Thomson expansion of Born-Infeld AdS Black Holes in consistent 4D Einstein-Gauss-Bonnet gravity}
\author{Chao-Ming Zhang$^{1}$\thanks{{\em
        email}: \href{mailto:zcm843395448@163.com}
        {zcm843395448@163.com}},
   Ming Zhang$^{2}$\thanks{{\em
        email}: \href{mailto:mingzhang0807@126.com}
        {mingzhang0807@126.com}},
and De-Cheng Zou$^{1}$ \thanks{Correspondence author {\em
        email}: \href{mailto:dczou@yzu.edu.cn}
        {dczou@yzu.edu.cn}}\\
$^{1}$Center for Gravitation and Cosmology, \\ College of Physical Science and Technology,\\
Yangzhou University, Yangzhou 225009, China\\
$^{2}$Faculty of Science, Xi'an Aeronautical University, Xi'an 710077 China}

\maketitle

\begin{abstract}
In this paper, the Joule-Thomson expansion of Born-Infeld AdS black holes in the consistent
Aoki-Gorji-Mukohyama theory of 4D Einstein-Gauss-Bonnet gravity is studied
in the extended phase space. We further analyze the effect of parameters $\alpha$ and $\beta$
on the inversion curves and plot the inversion and isenthalpic curves in the $T-P$ plane,
which can determine the cooling-heating regions.
\end{abstract}

\section{Introduction}\label{int}

\par
The Einstein's General Theory of Relativity (GR), the cornerstone of modern cosmology, provides an accurate description
of the phenomena in our universe. However, the observational data around black holes, with the strong gravity regime,
show the possibility of alternatives to the GR. To go beyond Einstein's theory, one usually modifies GR by adding some
higher derivative terms or new degrees of freedom of the metric to the field equations.
On the other hand, the string theory at the low energy limit also results in models of gravity with the action
including higher order curvature terms \cite{Gross:1986iv,Zwiebach:1985uq,Lust:1989tj}.
Recently, Glavan and Lin \cite{Glavan:2019inb} presented a non-trivial 4D Einstein-Gauss-Bonnet (EGB) gravity
by re-scaling the coupling constant $\alpha\rightarrow\alpha/(D-4)$ and taking the limit $D\rightarrow4$.
The 4D EGB gravity reports several appealing new predictions of this theory, including the singularity resolution
for spherically symmetric solutions.
However, the original claim of Ref. \cite{Glavan:2019inb} is clearly in contradiction with the Lovelock
theorem, and some subtleties and criticisms on the $D\rightarrow4$ limit \cite{Ai:2020peo,Mahapatra:2020rds,Shu:2020cjw,Tian:2020nzb,Arrechea:2020evj}. Very recently,
using the Arnowitt-Deser-Misner (ADM) decomposition, Aoki et al. \cite{Aoki:2020lig} proposed a
consistent description of this 4D theory that serves as a consistent
realization of $D\rightarrow4$ limit of the EGB gravity with two dynamical degrees of freedom
by breaking the temporal diffeomorphism invariance, and then a well-defined theory was formulated.
Along this line, the cosmological implications of this theory in the presence of a minimally coupled $k$-essence
field \cite{Aoki:2020iwm}, inflationary gravitational waves with investigation of the slow-roll single
field inflation \cite{Aoki:2020ila} and holographic superconductors in the 4D neutral AdS black
hole background \cite{Qiao:2020hkx} were further studied
in the consistent  Aoki-Gorji-Mukohyama theory of 4D EGB gravity.

On the other hand, black hole thermodynamics have been in the spotlight, since the discovery of black hole's
entropy \cite{Bekenstein:1973ur}, the four thermodynamic law \cite{Bardeen:1973gs}, and the
Hawking radiation of black hole \cite{Hawking:1975vcx} in 1970s. During these years,
the discussion of thermodynamics in AdS black holes has been generalized to the extended phase space,
where the cosmological constant is identified with thermodynamic pressure \cite{Dolan:2011xt,Dolan:2010ha}.
Later, a number of papers explored various thermodynamic aspects of black holes such as Van der waals like
small-large black hole phase transition \cite{Kubiznak:2012wp,Zou:2014mha}, reentrant phase transition \cite{Gunasekaran:2012dq,Zou:2013owa,Zou:2016sab}, heat engine's efficiency \cite{Johnson:2015fva,Johnson:2014yja},
compressibility \cite{Dolan:2011jm} in the extended phase space.
Apart from the phase transition and critical phenomena, the analogy has been done for the
Joule-Thomson (JT) expansion process in classical thermodynamics and black hole background \cite{Okcu:2016tgt}.
In classical thermodynamics, the JT expansion denotes a gas at a high pressure passes through a valve
or porous plug to a low pressure section such that during the process enthalpy is unchanged
and the process is an adiabatic expansion. Until now, the JT expansion of black holes has been
extensively studied in various backgrounds, such as Hayward AdS black hole \cite{Guo:2019gkr,Zhang:2021raw},
Bardeen AdS black hole \cite{Pu:2019bxf,Li:2019jcd}, charged AdS black holes in Rainbow gravity \cite{Yekta:2019wmt},
Kerr AdS black hole \cite{Okcu:2017qgo} and  and charged black hole in 4D EGB gravity \cite{Hegde:2020xlv}.

Recently, BI-AdS black hole solution was obtained in the consistent Aoki-Gorji-Mukohyama theory of
4D EGB gravity minimally coupled to BI electrodynamics \cite{Jafarzade:2020ova},
where Jafarzade et.al also recovered that this solution takes the same form of the solution
in the $D\rightarrow4$ limit of any D-dimensional solution of EGB gravity \cite{Yang:2020jno}.
In fact, as a well-known nonlinear electrodynamics theory, the BI electrodynamics was usually adopted to regularize
the ultraviolet divergent self-energy of a point-like charge in classical dynamics \cite{Born:1933pep}.
In Ref.~\cite{Zhang:2020obn}, we have further analyzed the critical behaviors of BI-AdS black hole
in 4D EGB gravity. Considering the increasing interest in study
of the consistent consistent Aoki-Gorji-Mukohyama theory of
4D EGB, JT expansion and BI electrodynamics, we here are going to investigate the JT Expansion
of BI-AdS black hole in the consistent Aoki-Gorji-Mukohyama theory of 4D EGB gravity.

The paper is organized as follows: we display the solution of BI-AdS black hole
and discuss the thermodynamics of black hole in the consistent 4D EGB gravity in Sect.\ref{the}.
In Sect.\ref{jou}, we discuss the JT expansion of BI-AdS black holes
in the consistent 4D EGB gravity, which include the JT coefficient, the inversion curves, the minimum inversion
temperature and the isenthalpic curves. We end the paper with closing remakes in the last section.

\section{Thermodynamics of Born-Infeld-AdS black hole} \label{the}

For the consistent Aoki-Gorji-Mukohyama theory of 4D EGB gravity, we can write the metric in the ADM formalism as
\begin{eqnarray}
ds^2=-N^2dt^2+\gamma_{ij}(dx^i+N^idt)(dx^j+N^jdt),
\end{eqnarray}
where $N$, $N^i$ and $\gamma_{ij}$ are the lapse function, the shift vector and spatial metric, respectively.
The well defined gravitational action in the presence of a negative
cosmological constant is given by \cite{Aoki:2020lig,Aoki:2020iwm,Aoki:2020ila}
\begin{eqnarray}\label{action}
{\cal S}_{g}=\int dt d^3x N\sqrt{\gamma}{\cal L}_{EGB}^{4D},
\end{eqnarray}
where the Lagrangian density
\begin{eqnarray}
{\cal L}_{EGB}^{4D}=&&\frac{M_{PI}^{2}}{2}\{2R+\frac{6}{L^2}-{\cal M}+\frac{\alpha}{2}[8R^2-4R{\cal M}-M^2\nonumber\\
&&-\frac{8}{3}(8R_{ij}R^{ij}-4R_{ij}{\cal M}^{ij}-{\cal M}_{ij}{\cal M}^{ij})]\},
\end{eqnarray}
with
\begin{eqnarray}
&&{\cal M}_{ij}\equiv R_{ij}+{\cal K}^{i}_{~i}{\cal K}_{ij}-{\cal K}_{ik}{\cal K}^{k}_{~j}, \quad
{\cal M}\equiv {\cal M}^{i}_{~i},\\
&&{\cal K}_{ij}\equiv \frac{1}{2N}[\dot{\gamma}_{ij}-2D_{(i} N_{j)}-\gamma_{ij}D^2 \lambda_{GF}],
\end{eqnarray}
Here $M^2_{PI}=(8pG)^{-1}$ is the reduced Planck mass, $R$ and $R_{ij}$ are respectively the Ricci
scalar and Ricci tensor of the spatial metric, and the factor $\lambda_{GF}$ is the
Lagrange multiplier related to a gauge-fixing constraint, see Ref.\cite{Aoki:2020lig}.
Moreover,  the dot denotes the derivative with respect to the time $t$, and $D_i$ represents the
covariant derivative compatible with the spatial metric.

In this paper, we plan to discuss JT expansion of BI-AdS black hole in
the consistent Aoki-Gorji-Mukohyama theory of 4D EGB gravity.
The gravitational Lagrangian  coupled to the BI electrodynamics Lagrangian is
\begin{eqnarray}
{\cal L}_{BI}=4\beta^2\left(1-\sqrt{1+\frac{F_{\mu\nu}F^{\mu\nu}}{2\beta^2}}\right), \quad
F_{\mu\nu}=\partial_{\mu}A_{\nu}-\partial_{\nu}A_{\mu},
\end{eqnarray}
where $\beta>0$ denotes the BI parameter which is the maximum of the electromagnetic field
strength.

Considering the 4-dimensional topological black hole, we take a static spherically symmetric metric ansatz
\begin{eqnarray}
ds^2=-f(r)dt^2+\frac{1}{f(r)}dr^2+r^2d\Omega_2^2,
\end{eqnarray}
where $d\Omega_2^2$ represents the metirc of a 2-dimensional unit sphere and the metric function $f(r)$
takes the form \cite{Jafarzade:2020ova,Yang:2020jno}
\begin{eqnarray}
f(r)&=&1+\frac{r^2}{2\alpha}\Big[1-\left(1+4\alpha\left(\frac{2M}{r^3}-\frac{1}{l^2}
-\frac{2\beta^2}{3}+\frac{2\beta^2}{3}\sqrt{1+\frac{Q^2}{\beta^2r^4}}\right.\right.\nonumber\\
&&\left.\left.-\frac{4Q^2}{3r^4}~_{2}F_1[\frac{1}{4},\frac{1}{2},\frac{5}{4},-\frac{Q^2}{\beta^2 r^4}]\right)\right)^{1/2}\Big].\label{solution}
\end{eqnarray}
Here $_{2}F_1$ is the hypergeometric function, the parameters $Q$ and $M$ are related to the black
holes charge and mass respectively. In Ref.~\cite{Jafarzade:2020ova}, it has also shown
that this solution (Eq.(\ref{solution})) is a solution of the
consistent Aoki-Gorji-Mukohyama theory (Eq.(\ref{action})).
In the limit of $\beta\rightarrow\infty$, $f(r)$ recovers the
RN-AdS-like black hole solution in the 4D EGB gravity.
In addition, $f(r)$ recovers the BI-AdS black hole solution obtained in the
Einstein-Born-Infeld gravity in the limit $\alpha\to 0$.

From the black hole solution $f(r)$, we can obtain the mass $M$,
Hawking temperature $T$
\begin{eqnarray}
M&=&\frac{r_+}{2}\Big[1+\frac{r_+^2}{l^2}+\frac{\alpha}{r_+^2}
+\frac{2\beta^2r_+^2}{3}\left(1-\sqrt{1+\frac{Q^2}{\beta^2r_+^4}}\right) \nonumber\\
&+&\frac{4Q^2}{3r_+^2}~_{2}F_1[\frac{1}{4},\frac{1}{2},\frac{5}{4},-\frac{Q^2}{\beta^2 r_+^4}]\Big],\label{M}\\
T&=&\frac{\Big[3r_+^4+l^2\left(r_+^2-\alpha +2\beta^2r_+^4\left(1-\sqrt{1+\frac{Q^2}{\beta^2r_+^4}}\right)\right)\Big]}{4\pi r_+l^2(r_+^2+2\alpha)}.\label{T}
\end{eqnarray}
On the other hand, the Bekenstein-Hawking entropy-area law is modified by the GB coupling parameter $\alpha$
with \cite{Fernandes:2020rpa,Wei:2020poh}
\begin{eqnarray}
S=\pi r_+^2+2\pi\alpha\ln{\left(\frac{r_+^2}{\alpha}\right)}.\label{S}
\end{eqnarray}

In the extended phase space, the relationship between the cosmological constant $\Lambda$ and
pressure $P$ can be expressed as $P=-\frac{\Lambda}{8\pi}=\frac{3}{8\pi l^2}$\cite{Dolan:2011xt,Dolan:2010ha}.
In the Born-Infeld case, $M$ should be the function of entropy $S$, pressure $P$, charge $Q$,
GB coupling parameter $\alpha$ and BI coupling coefficient $\beta$. Moreover, those thermodynamic
quantities satisfy the following differential form
\begin{eqnarray}
dM=TdS+\Phi dQ+{\cal B}d\beta+VdP+{\cal A} d\alpha,
\end{eqnarray}
where the thermodynamic volume $V$ conjugate to $P$ equals to
\begin{eqnarray}\label{V}
V=\frac{4\pi r_+^3}{3}
\end{eqnarray}
and $\Phi$ is the electromagnetic potential difference between the horizon and infinity
\begin{eqnarray}
\Phi=\frac{Q}{r}{_2}F_1[\frac{1}{4},\frac{1}{2},\frac{5}{4},-\frac{Q^2}{\beta^2r_+^{4}}].
\end{eqnarray}
Moveover, ${\cal B}$ and ${\cal A}$ are the conjugate quantities of BI coupling parameter $\beta$
and Gauss-Bonnet coupling $\alpha$ respectively
\begin{eqnarray}
&&{\cal B}=\frac{2\beta r_+^{3}}{3}\left(1-\sqrt{1+\frac{Q^2}{\beta^2r_+^{4}}}\right)
+\frac{Q^2}{3\beta r_+}{_2}F_1[\frac{1}{4},\frac{1}{2},\frac{5}{4},-\frac{Q^2}{\beta^2r_+^{4}}],\\
&&{\cal A}=\frac{1}{2r_+}+2\pi T-2\pi T\ln{\left(\frac{r_+^2}{\alpha}\right)}.
\end{eqnarray}
By scaling argument, we can obtain the generalized Smarr
relation for the BI-AdS black holes in the
extended phase space
\begin{eqnarray}
M=2T S+\Phi Q-{\cal B} \beta-2VP+2{\cal A} \alpha.
\end{eqnarray}

In addition, we can get the equation of state for the black hole system
\begin{eqnarray}\label{eqstate}
&&P=-\frac{\beta^2}{4\pi}+\frac{\beta^2\sqrt{1+\frac{Q^2}{\beta^2r_+^4}}}{4\pi}-\frac{1}{8\pi r_+^2}
+\frac{T}{2r_+}+\frac{\alpha}{8\pi r_+^4}+\frac{T\alpha}{r_+^3}
\end{eqnarray}
We know that the critical points occur when $P$ has an
inflection point \cite{Kubiznak:2012wp}
\begin{eqnarray}
\frac{\partial P}{\partial r_+}\Big|_{T=T_c, r_+=r_c}
=\frac{\partial^2 P}{\partial r_+^2}\Big|_{T=T_c, r_+=r_c}=0.\label{inflection}
\end{eqnarray}
Then, we can obtain critical temperature and critical pressure
\begin{eqnarray}
T_c=&&\frac{r_c^2-2\alpha-2Q^2(1+\frac{Q^2}{\beta^2r_c^4})^{-1/2}}{2\pi r_c(r_c^2+6\alpha)},\label{critTc}\\
P_c=&&\frac{r_c^4-5\alpha r_c^2-2\alpha^2}{8\pi r_c^4(r_c^2+6\alpha)}
+\frac{\beta^2}{4\pi\sqrt{1+\frac{Q^2}{\beta^2r_c^4}}}-\frac{\beta^2}{4\pi} \nonumber\\
&&-\frac{Q^2(r_c^2+2\alpha)}{4\pi r_c^4(r_c^2+6\alpha)\sqrt{1+\frac{Q^2}{\beta^2r_c^4}}}.\label{critPc}
\end{eqnarray}

\section{Joule Thomson Expansion}\label{jou}

In the Joule-Thomson expansion process, increasing pressure slowly pushes the gas through the plug
under a sealed and insulated condition, and finally achieves equilibrium on both sides. We can use
the Joule-Thomson(JT) coefficient to describe the ratio
of the temperature change to the pressure change after the expansion of a gas
\begin{eqnarray}
&&\mu=\left(\frac{\partial T}{\partial P}\right)_H=\frac{1}{C_P}\Big[T\left(\frac{\partial V}{\partial T}\right)_P-V\Big].
\end{eqnarray}
This coefficient $\mu$ characterizes the expansion and plays an important role as its sign describes
whether the heat is absorbed or evolved during the expansion process. It is clear that the
system will experience a cooling (heating) process with $\mu>0(\mu<0)$. Considering $\mu=0$, we can obtain
the inversion point $(T_i,P_i)$ with
\begin{eqnarray}\label{inverTi}
T_i=V\left(\frac{\partial T}{\partial V}\right)_{P_i}.
\end{eqnarray}

Recently, Shihao Bi \cite{Bi:2020vcg} studied the JT expansion of 4D BI-AdS black hole in general relativity.
It's worthy to point out that there is only one inversion curve
in the 4D BI-AdS black hole system, since black hole system usually has only one inversion temperature,
corresponding to the lower inversion temperature of Van der Waals fluid \cite{Bi:2020vcg}.
For Van der Waals Fluids system, the inversion temperature has two branches ($T_i^{lower}$ and $T_i^{upper}$),
see Ref.~\cite{Okcu:2016tgt}.

\begin{figure}
\begin{minipage}[t]{0.5\linewidth}
\includegraphics[width=5.5cm,height=3.4cm]{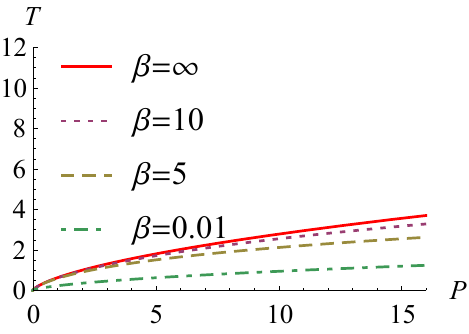}
\hfill%
\end{minipage}
\begin{minipage}[t]{0.5\linewidth}
\includegraphics[width=5.5cm,height=3.4cm]{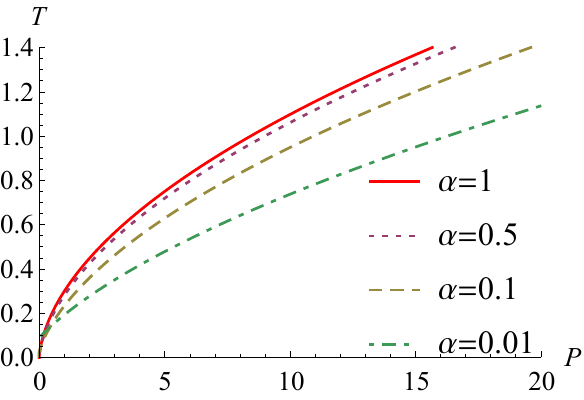}
\end{minipage}
\caption{Inversion curves for Born-Infeld AdS Black Holes in $T_i-P_i$ plane with $Q=1$.
In the left panel, we take $\alpha=0.1$, $\beta=0.01, 5, 10, \infty$. In the right panel,
we take $\beta=0.1$,$\alpha=0.01,0.1,0.5,1$.}\label{figtp}
\end{figure}

Here we turn to consider JT expansion of BI-AdS Black Hole in consistent 4D EGB gravity.
Substituting Eqs.(\ref{T})(\ref{V}) into Eq.(\ref{inverTi}), we can obtain
\begin{eqnarray}\label{ti}
T_i&=&\frac{2\beta^2\left(r_{+i}^6+6\alpha r_{+i}^4\right)+8\pi P_i\left(r_{+i}^6+6\alpha r_{+i}^4\right)
-r_{+i}^4+2\alpha^2+5\alpha r_{+i}^2}{12\pi r_{+i}\left(2\alpha+r_{+i}^2\right)^2}\nonumber\\
&&+\frac{Q^2\left(r_{+i}^2-2\alpha\right)-\beta^2 \left(r_{+i}^6+6 \alpha r_{+i}^4\right)}{6 \pi r_{+i}\left(2 \alpha
+r_{+i}^2\right)^2\sqrt{\frac{Q^2}{\beta^2 r_{+i}^4}+1}}.\label{ti}
\end{eqnarray}
with $P=\frac{3}{8\pi l^2}$.
In addition, the inverse temperature $T_i$ and pressure $P_i$ also satisfy the equation of state (Eq.(\ref{eqstate})).
Considering the equation of state, we can obtain the inversion temperature $T_i$ and pressure $P_i$
\begin{eqnarray}
P_i&=&\frac{\beta^2 r_{+i}^6+2Q^2\left(\alpha+r_{+i}^2\right)}{4\pi r_{+i}^4\sqrt{\frac{Q^2}{\beta^2}+r_{+i}^4}}
+\frac{-2\beta^2 r_{+i}^6-2r_{+i}^4+\alpha r_{+i}^2+4\alpha^2}{8\pi r_{+i}^6},\label{Pi}\\
T_i&=&\frac{Q^2}{2\pi r_{+i}\sqrt{\frac{Q^2}{\beta^2}+r_{+i}^4}}-\frac{r_{+i}^2-2\alpha}{4\pi r_{+i}^3}.\label{Ti}
\end{eqnarray}
Via Eqs.~(\ref{Pi}) and (\ref{Ti}), the inversion curves of black hole are plotted in Fig.~\ref{figtp}
for different values of parameters $\alpha$, $\beta$ and fixed $Q$. We can find the inversion temperature increases monotonously with the inversion pressure.
For $\alpha=0.1$, the left panel of Fig.~\ref{figtp} shows the effect of $\beta$ on the inversion curves. With the increasing of $\beta$, the inversion temperature for given pressure tends to increase, and the slope gradually decreases. This is consistent with the results obtained by Shihao Bi \cite{Bi:2020vcg} in the limit of $\beta\rightarrow\infty$.
By fixing $\beta=0.1$, the right panel of Fig.~\ref{figtp} exhibits the effect of $\alpha$ on the inversion curves. The inversion temperature also increases with the increasing of $\alpha$.
As mentioned above, the inversion curve of BI-AdS black hole is not closed, which means the black holes always cool above the inversion curve during the Joule-Thomson expansion.

\begin{figure}
\centering
\includegraphics[width=8cm,height=5.5cm]{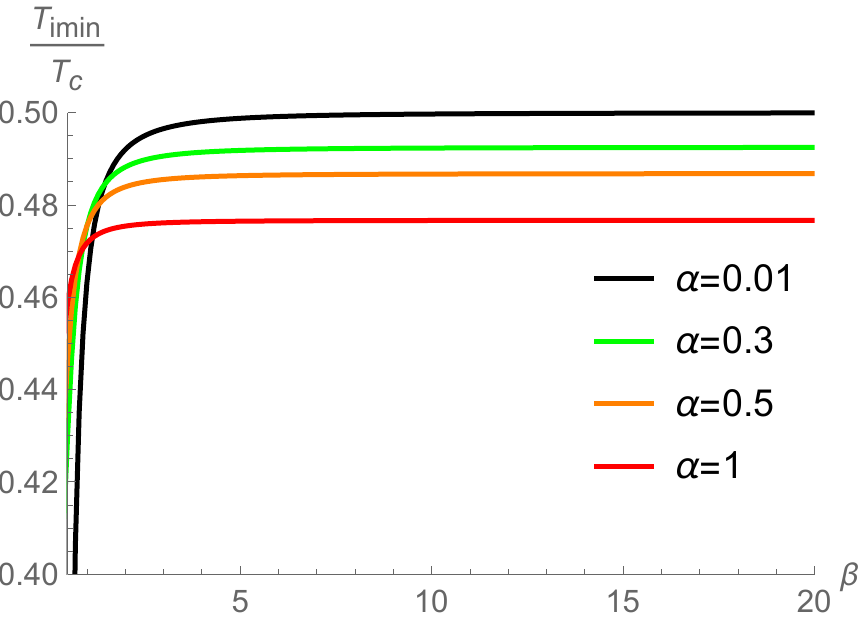}
\caption{The ratio of the minimum inversion temperature to the critical temperature, we take $Q=1$ and $\alpha=0.01,0.05,0.1,0.5,1.$}\label{TITC}
\end{figure}

\begin{figure}[htb]
\subfigure[$\alpha=0$]{\label{titca} 
\includegraphics[width=2.3in]{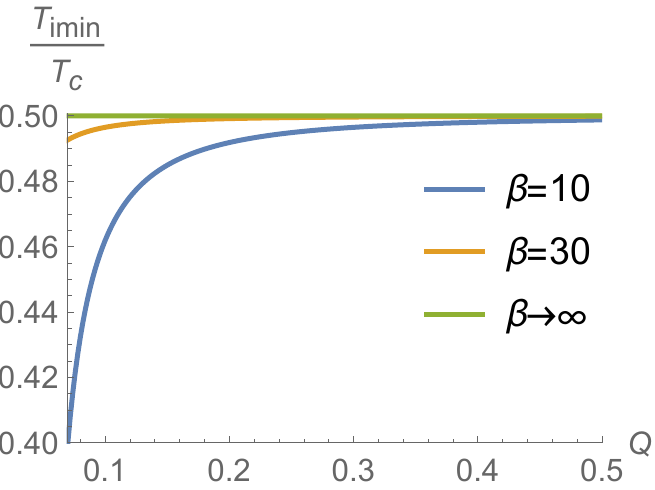}}
\hfill%
\subfigure[$\alpha=0.02$]{\label{titcb}
\includegraphics[width=2.3in]{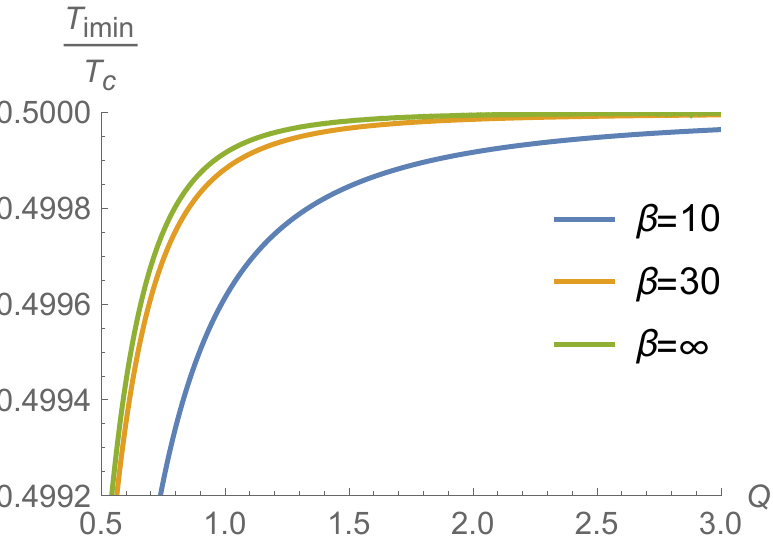}}
\caption{Ratio of minimum inversion temperature to critical temperature. We take $\alpha=0,0.02$, in each diagram, $\beta=10,30,\infty$.}\label{figtitc}
\end{figure}

\begin{figure}[htb]
\subfigure[$\beta=10$]{\label{titcc} 
\includegraphics[width=2.3in]{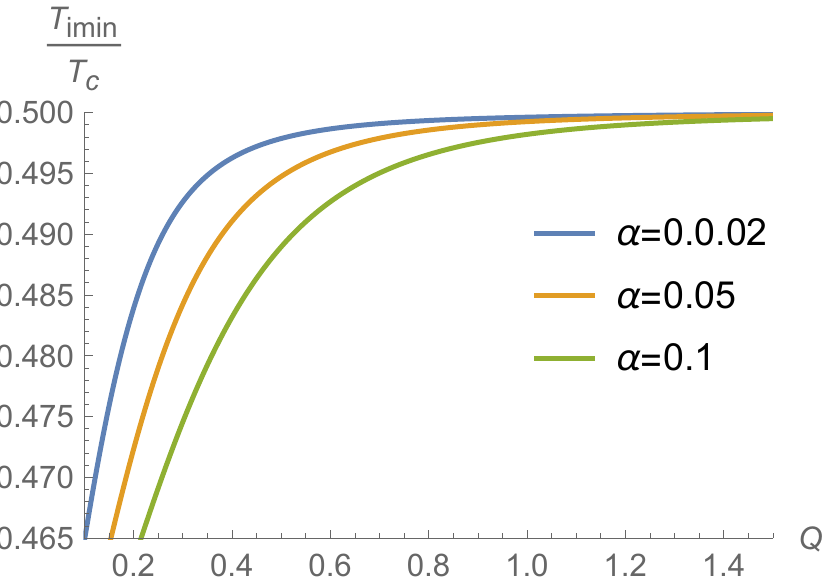}}
\hfill%
\subfigure[$\beta \to \infty$]{\label{titcd}
\includegraphics[width=2.3in]{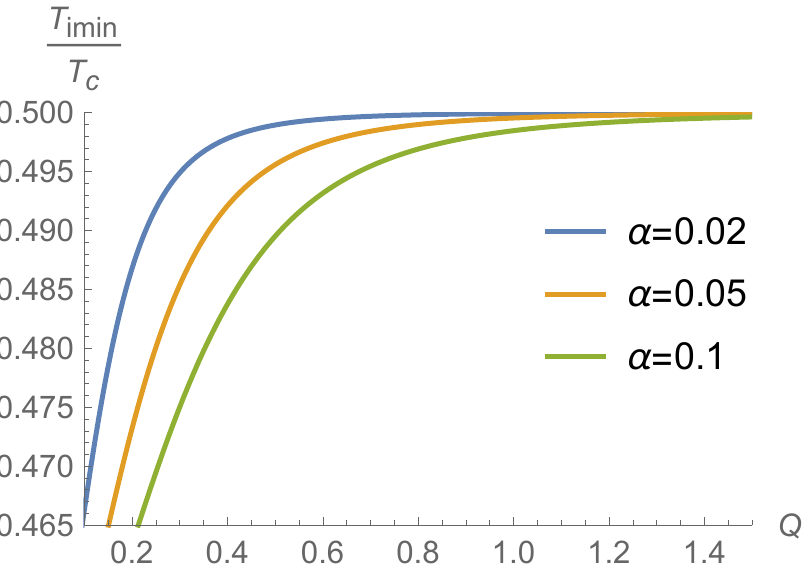}}
\caption{Ratio of minimum inversion temperature to critical temperature. We take $\beta=10,\infty$,
in each diagram $\alpha=0.02,0.05,0.1$ .}\label{figtitc1}
\end{figure}

The minimum inversion temperature $T_i^{\text{min}}$ occurs at the point $P_i=0$. Since there are higher order terms in $P_i$, the minimum
inversion temperature $T_i^{\text{min}}$ can be obtained numerically. It has been shown that the ratio of minimum
inversion temperature to the critical temperature $\frac{T_i^{\text{min}}}{T_c}$ for RN-AdS black holes in GR equals to 1/2 in Ref.\cite{Okcu:2016tgt},
but here we show this value will correct in the
case of parameters $\beta$ and $\alpha$. We show the $\beta$ dependence of the ratio $\frac{T_i^{\text{min}}}{T_c}$ with different $\alpha$ in Fig.\ref{TITC},
and the $Q$ dependence of the ratio $\frac{T_i^{\text{min}}}{T_c}$ with different $\beta$ in Fig.\ref{figtitc} and different $\alpha$ in Fig.\ref{figtitc1}.
We can find that the ratio $\frac{T_i^{\text{min}}}{T_c}$ is not always $\frac{1}{2}$, but the curves show that the ratio $\frac{T_i^{\text{min}}}{T_c}$ tends to $\frac{1}{2}$ as $Q$ or $\beta$ increases. When $\alpha=0$ and $\beta\to \infty$, it degenerates into RN AdS black hole and the ratio $\frac{T_i^{\text{min}}}{T_c}$ always $\frac{1}{2}$.

\begin{figure}
\begin{minipage}[t]{0.5\linewidth}
\includegraphics[width=6.5cm,height=5cm]{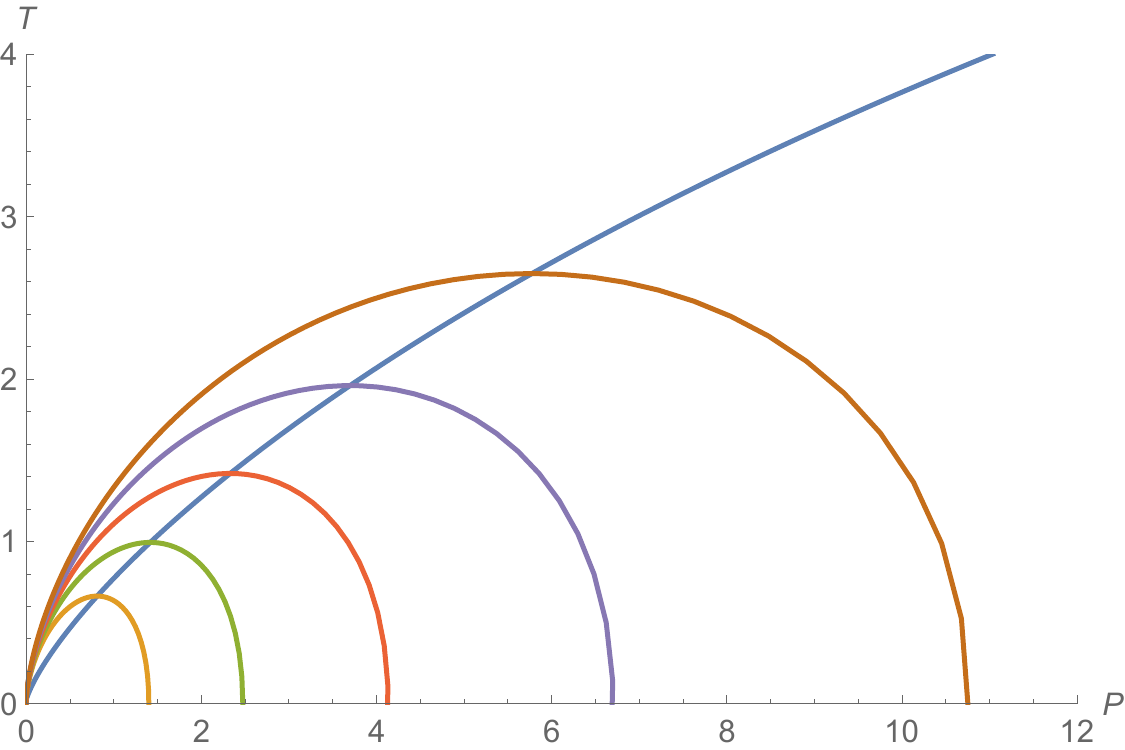}
\centering (a)$Q=1,\alpha=0,\beta=10$
\hfill%
\label{Ha}
\end{minipage}%
\begin{minipage}[t]{0.5\linewidth}
\includegraphics[width=6.5cm,height=5cm]{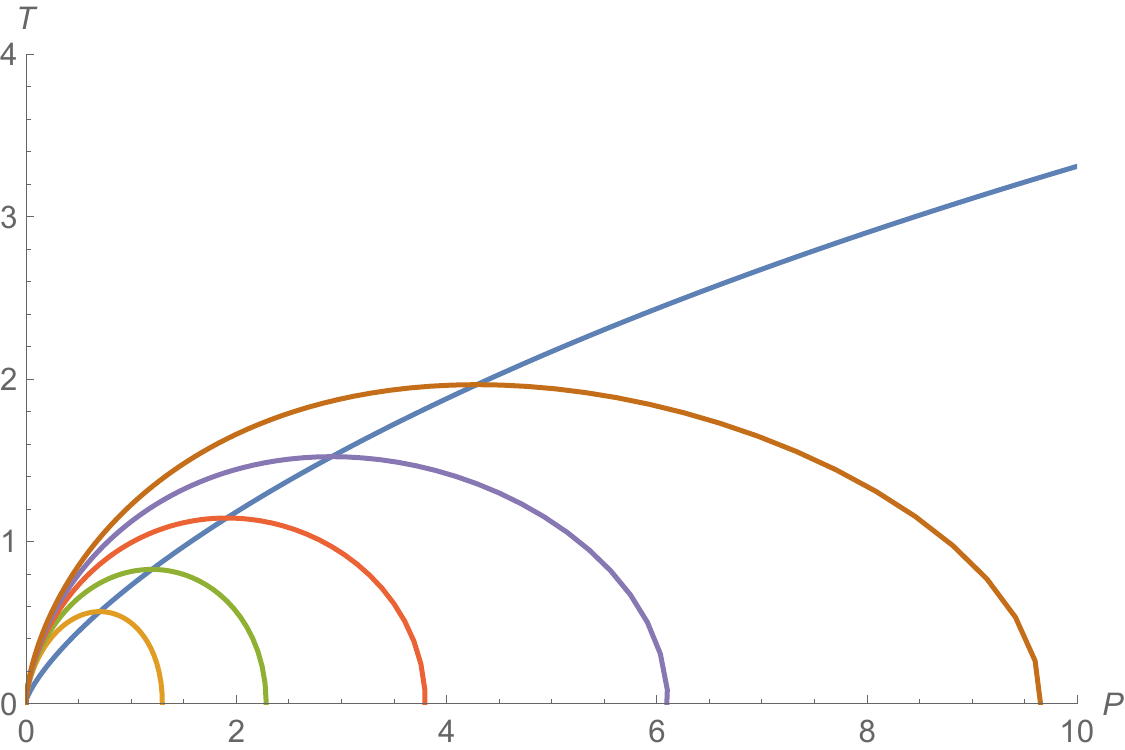}
\centering (b)$Q=1,\alpha=0.02,\beta=10$
\hfill%
\label{Hb}
\end{minipage}
\par
\begin{minipage}[t]{0.5\linewidth}
\includegraphics[width=6.5cm,height=5cm]{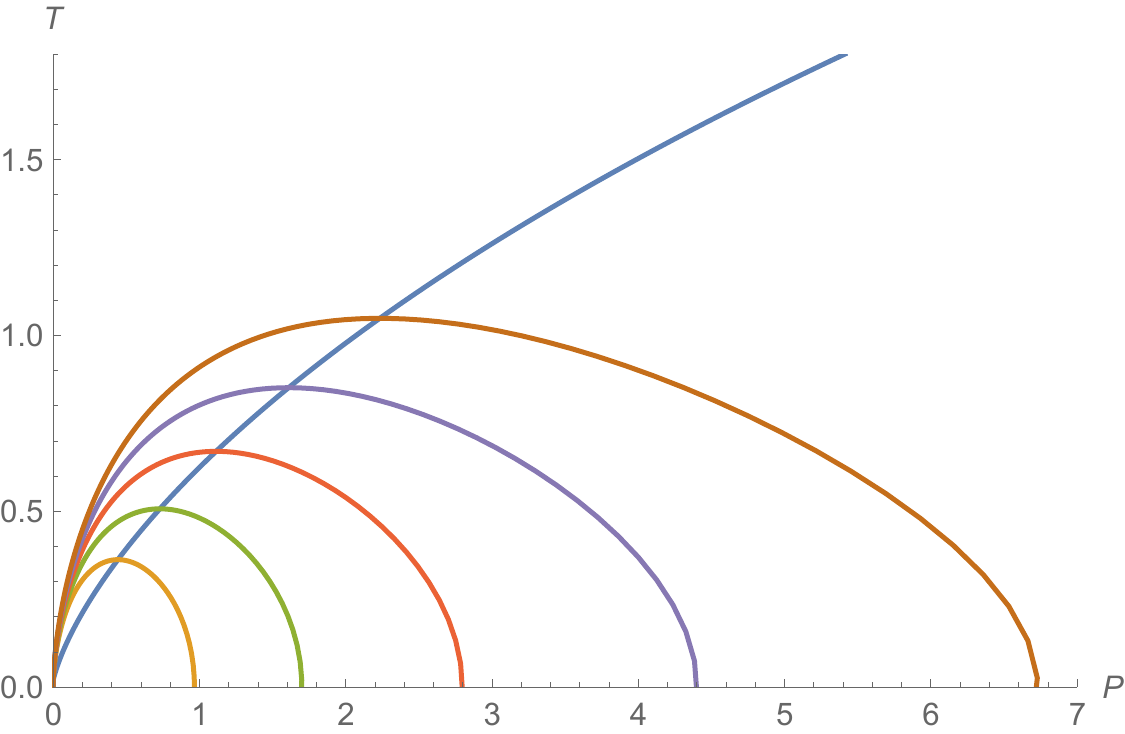}
\centering (c)$Q=1,\alpha=0.1,\beta=10$
\hfill%
\label{Hc}
\end{minipage}%
\begin{minipage}[t]{0.5\linewidth}
\includegraphics[width=6.5cm,height=5cm]{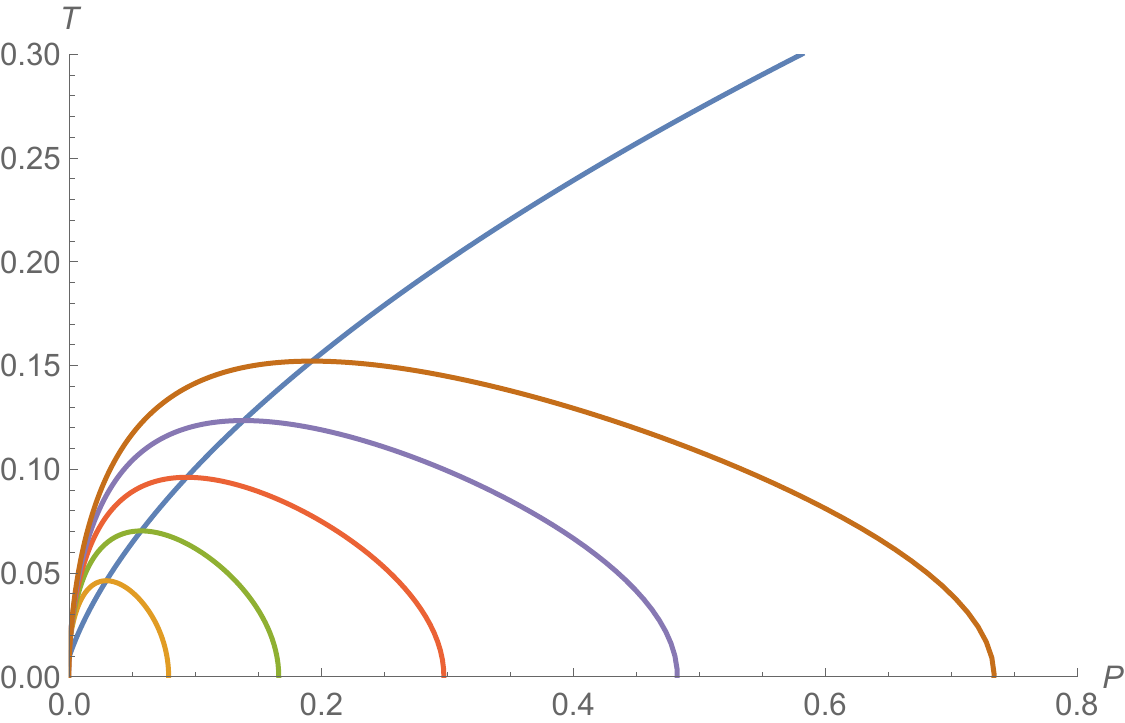}
\centering (d)$Q=1,\alpha=1,\beta=10$
\hfill%
\label{Hd}
\end{minipage}%
\caption{ Inversion curve and isenthalpic curves for Born-Infeld AdS Black Holes in 4D EGB gravity.
The line through closed curves is the inversion curve, the closed curves are isoenthalpy curves.
Here we take $M=1.8,2.0,2.2,2.4,2.6.$}
\label{H}
\end{figure}

On the other hand, mass $M$ can be interpreted as enthalpy in the extended phase space \cite{Dolan:2011xt,Dolan:2010ha,Kubiznak:2012wp}.
Considering Eqs.(\ref{M})(\ref{T}) with $P=\frac{3}{8\pi l^2}$, the pressure $P$ and temperature $T$ can be rewritten
as functions of $M$ and $r_+$.
\begin{eqnarray}
P=&&\frac{1}{8\pi r_+^4}(6M r_+ -3r_+^2-2\beta^2 r_+^4 +2\beta r_+^4 \sqrt{\frac{Q^2+\beta^2 r_+^4}{\beta^2 r_+^4}} \nonumber\\ &&-3\alpha-4Q^2~_{2}F_1[\frac{1}{4},\frac{1}{2},\frac{5}{4},-\frac{Q^2}{\beta^2 r^4}]),\label{PJ}\\
T=&&\frac{-3M r_+ +r_+^2+2\alpha +2Q^2~_{2}F_1[\frac{1}{4},\frac{1}{2},\frac{5}{4},-\frac{Q^2}{\beta^2 r^4}]}{2\pi r_+^3 +4\pi r_+ \alpha}. \label{TJ}
\end{eqnarray}
Then, we can plot the isoenthalpy curve in the $T-P$ plane by fixing the mass $M$ of BI-AdS black hole.
The isoenthalpy curves and the inversion curves of BI-AdS black holes in 4D EGB
gravity are shown in Fig.\ref{H}, it shows the inversion curve is the dividing line between heating 
and cooling process. Note that the isoenthalpy curve intersects the inversion curve at the inversion 
point which also is the maximum point for a specific isenthalpic curve, representing at the inversion 
point the temperature is highest during the whole Joule-Thomson expansion process. Above the inversion curve, 
the slope of the isenthalpic curve is positive, there is a cooling process. On the contrary, 
the slope changes to negative and the heating occurs below the inversion curve.
Meanwhile, the isoenthalpy curve tends to shrink with increase of GB coupling constant $\alpha$.

\section{Final Remarks}\label{con}

In the 4-dimensional Einstein Gauss-Bonnet gravity, we have studied the Joule-Thomeson expansion of BI-AdS black hole in the extended phase space.
We obtained the thermodynamic quantities and the first law of black hole thermodynamic. Then, the well-known JT coefficient $\mu$ was derived via the first law of black hole thermodynamics. The zero point of $\mu$ is the inversion point which discriminate the cooling process from heating process.

We studied the dependence of $\alpha$ and $\beta$ on the inversion curves, the results were depicted in Fig. \ref{figtp}. It is significant to calculate the ratio between the minimum inversion temperature $T_i^{\text{min}}$ and the critical temperature $T_c$. Fig. \ref{figtitc} and Fig. \ref{figtitc1} displayed the ratio versus the charge $Q$, parameters $\alpha$ and $\beta$. The ratio tends to $\frac{1}{2}$ as $Q$ or $\beta$ increases. We also plot the isenthalpic curves and the inversion curves in Fig.\ref{H}, which shows the slope of the inversion curve is always positive. This result means the black hole
always cools (heats) above (below) the inversion curve during the expansion. For different values of $\alpha$, $Q$ and $\beta$,
we can distinguish the cooling and heating regions with the inversion curve.

\section*{Acknowledgements}

This work is supported by outstanding young teacher programme from Yangzhou University, No. 137050368.

\end{document}